\documentclass[11pt,a4paper]{article}
\pdfoutput=1

\usepackage{jheppub}
\usepackage{latexsym}
\usepackage{revsymb}
\usepackage{multirow}
\usepackage{color}
\usepackage[usenames,dvipsnames,svgnames,table]{xcolor}

\usepackage{graphicx}
\usepackage{epsfig}  
\usepackage{epsf}    
\usepackage{dcolumn}
\usepackage{bm}
\usepackage{dcolumn}
\usepackage{textcomp}
\usepackage{float}
\usepackage{hypcap}
\usepackage[]{hyperref}
\usepackage{adjustbox}
\usepackage{multirow}

\usepackage{subfigure}
\usepackage{alphalph}
\usepackage{morefloats}
\usepackage{textcomp}

\newcommand{\be}{\begin{equation}}
\newcommand{\ee}{\end{equation}}
\newcommand{\ba}{\begin{eqnarray}}
\newcommand{\ea}{\end{eqnarray}}

\preprint{IP/BBSR/2017-16}

\title{Signatures of a Light Sterile Neutrino in T2HK} 

\author[a,b,c]{Sanjib Kumar Agarwalla,}
\author[a,b]{Sabya Sachi Chatterjee,}
\author[d,e]{Antonio Palazzo}

\affiliation[a]{Institute of Physics, Sachivalaya Marg, Sainik School Post, Bhubaneswar 751005, India}
\affiliation[b]{Homi Bhabha National Institute, Training School Complex, Anushakti Nagar, Mumbai 400085, India}
\affiliation[c]{International Centre for Theoretical Physics, Strada Costiera 11, Trieste 34151, Italy}
\affiliation[d]{Dipartimento Interateneo di Fisica ``Michelangelo Merlin'', Via Amendola 173, 70126 Bari, Italy} 
\affiliation[e]{Istituto Nazionale di Fisica Nucleare (INFN), Sezione di Bari, Via E.\ Orabona 4, I-70126 Bari, Italy}

\emailAdd{sanjib@iopb.res.in}
\emailAdd{sabya@iopb.res.in}
\emailAdd{palazzo@ba.infn.it}

\abstract{
We investigate the performance of T2HK in the presence of a light eV scale sterile neutrino.
We study in detail its influence in resolving fundamental issues like mass hierarchy,
CP-violation (CPV) induced by the standard CP-phase $\delta_{13}$ and new CP-phase $\delta_{14}$,
and the octant ambiguity of $\theta_{23}$. We show for the first time in detail that due to the 
impressive energy reconstruction capabilities of T2HK, the available spectral information plays 
an important role to enhance the mass hierarchy discovery reach of this experiment 
in 3$\nu$ framework and also to keep it almost intact even in $4\nu$ scheme. 
This feature is also of the utmost importance in establishing the CPV due to $\delta_{14}$. As far as 
the sensitivity to CPV due to $\delta_{13}$ is concerned, it does not change much 
going from $3\nu$ to 4$\nu$ case. We also examine the
reconstruction capability of the two phases $\delta_{13}$ and $\delta_{14}$, and find that
the typical 1$\sigma$ uncertainty on $\delta_{13}$ ($\delta_{14}$) in T2HK is 
$\sim15^0$ ($30^0$). While determining the octant of $\theta_{23}$, we face a complete 
loss of sensitivity for unfavorable combinations of unknown $\delta_{13}$ and $\delta_{14}$.}

\arxivnumber{1801.04855}
\begin{document}
\maketitle

\section{Introduction and Motivation}
\label{introduction}

Sterile neutrinos, neutral singlets of the SU(2)$_L$ weak isospin group, are perhaps the simplest extension of the Standard Model.
Intriguingly, a series of anomalies detected at the short baseline (SBL) experiments, 
support the existence of one or more sterile states (see~\cite{Abazajian:2012ys,Palazzo:2013me,Gariazzo:2015rra} 
for a review of the topic) at the scale of $\sim$ 1\,eV. 

New SBL experiments will be able to test such a hypothesis (see the review in~\cite{Lasserre:2014ita})
seeking the typical $L/E$ pattern induced by the oscillations  driven by the 
new mass-squared splitting(s) $\Delta m^2_{new} \sim 1$\,eV$^2$. However, 
after a hypothetical discovery, the complete mapping of the parameters regulating
the oscillation process would require other kinds of experiments. In fact, even in the simplest scenario, 
the so called 3+1 scheme, where only one sterile state participates to the oscillations,
there are three new mixing angles and two new CP violating phases in addition to
the standard 3-flavor parameters.

The CP phases (both standard and non-standard) can be observed solely through 
interference effects between two distinct oscillation frequencies. In the 3+1 scheme,
at SBL setups, only one frequency is observable (the new one) while the atmospheric and
solar frequencies are basically invisible. This drastically suppresses the amplitude of
any interference effect, and, as a consequence, the SBL experiments 
are completely blind to the CP violation phases.

As first shown in~\cite{Klop:2014ima}, the situation is qualitatively different at the long baseline (LBL) experiments, 
because in these setups the interference between two distinct frequencies shows up. In this
way the LBL experiments acquire sensitivity both to the standard CP phase and to
the new ones. Therefore, the LBL setups have a complementary role to the SBL experiments
in the searches for the sterile neutrino properties. 

The next-generation LBL neutrino oscillation experiments~\cite{Pascoli:2013wca,Agarwalla:2013hma,Agarwalla:2014fva,Feldman:2012qt,Stanco:2015ejj}, would naturally play a crucial role in this arena. 
In the present work, we focus on the proposed experiment 
Tokai to Hyper Kamiokande (T2HK), which will make use of a very powerful neutrino beam shot 
from Tokai to Kamioka over a distance of 295 km. Our work complements other recent studies performed 
about DUNE~\cite{Hollander:2014iha,Berryman:2015nua,Gandhi:2015xza,Coloma:2017ptb,Tang:2017khg} and 
T2HK~\cite{Kelly:2017kch,Choubey:2017cba,Choubey:2017ppj,Tang:2017khg}.
Previous studies on sterile neutrinos at LBL experiments can be found 
in~\cite{Donini:2001xy,Donini:2001xp,Donini:2007yf,Dighe:2007uf,Donini:2008wz,Yasuda:2010rj,Meloni:2010zr,Bhattacharya:2011ee,Donini:2012tt}. 

The paper is organized as follows. In section~\ref{sec:probability}, we present the theoretical
framework and discuss the behavior of the 4-flavor $\nu_\mu \to \nu_e$ 
conversion probability.  In section~\ref{sec:T2HK}, we describe the details 
of the T2HK setup and of the numerical analysis. In section~\ref{sec:MH},
we present the sensitivity study of the mass hierarchy. 
Section~\ref{sec:CPV} deals with the CPV discovery potential 
and the reconstruction capability of the CP phases. Section~\ref{sec:octant} is
devoted to the assessment of the capability of reconstructing the octant of $\theta_{23}$.
 Finally, we draw our conclusions in section~\ref{Conclusions}. 

\section{Transition probability in the 3+1 scheme}
\label{sec:probability}
\subsection{Theoretical framework}

In the 3+1 scheme, the four flavor eigenstates ($\nu_e,\nu_\mu,\nu_\tau, \nu_s$)
are related to the mass eigenstates ($\nu_1,\nu_2,\nu_3,\nu_4$) through a $4\times4$ 
unitary matrix, which can be written in the following fashion
\begin{equation}
\label{eq:U}
U =   \tilde R_{34}  R_{24} \tilde R_{14} R_{23} \tilde R_{13} R_{12}\,, 
\end{equation} 
where $R_{ij}$ ($\tilde R_{ij}$) is a real (complex) $4\times4$ rotation 
matrix with a mixing angle $\theta_{ij}$ and containing the $2\times2$ 
submatrix 
\begin{eqnarray}
\label{eq:R_ij_2dim}
     R^{2\times2}_{ij} =
    \begin{pmatrix}
         c_{ij} &  s_{ij}  \\
         - s_{ij}  &  c_{ij}
    \end{pmatrix} ,
\,\,\,\,\,\,\,   
     \tilde R^{2\times2}_{ij} =
    \begin{pmatrix}
         c_{ij} &  \tilde s_{ij}  \\
         - \tilde s_{ij}^*  &  c_{ij}
    \end{pmatrix}
\,,    
\end{eqnarray}
in the  $(i,j)$ sub-block with
\begin{eqnarray}
 c_{ij} \equiv \cos \theta_{ij}, \qquad s_{ij} \equiv \sin \theta_{ij}, \qquad  \tilde s_{ij} \equiv s_{ij} e^{-i\delta_{ij}}.
\end{eqnarray}
The parameterization adopted in Eq.~(\ref{eq:U}) has several merits: i) In the limiting
case in which there is no mixing of the active states with the sterile one 
$(\theta_{14} = \theta_{24} = \theta_{34} =0)$, 
it gives back the 3-flavor matrix in its usual parameterization.
ii) For small values of $\theta_{13}$, $\theta_{14}$, and $\theta_{24}$, 
one has $|U_{e3}|^2 \simeq s^2_{13}$, $|U_{e4}|^2 = s^2_{14}$, 
$|U_{\mu4}|^2  \simeq s^2_{24}$, and $|U_{\tau4}|^2 \simeq s^2_{34}$, 
with a transparent physical interpretation of the new mixing angles. 
iii) The leftmost positioning of the rotation matrix $\tilde R_{34}$ makes 
the vacuum $\nu_{\mu} \to \nu_{e}$ transition probability independent of $\theta_{34}$
and of the associated CP phase $\delta_{34}$ (see~\cite{Klop:2014ima}). 
\subsection{Transition probability}
\label{subsec:vacuum}

For the T2HK baseline (295 km), matter effects are not significant.
Therefore, we can restrict our analytical discussion to the vacuum case. 
As discussed in~\cite{Klop:2014ima}, the $\nu_{\mu} \to \nu_{e}$ 
transition probability can be expressed as the sum of three terms
\begin{eqnarray}
\label{eq:Pme_4nu_3_terms}
P^{4\nu}_{\mu e}  \simeq  P^{\rm{ATM}} + P^{\rm {INT}}_{\rm I}+   P^{\rm {INT}}_{\rm II}\,.
\end{eqnarray}
The first term is controlled by the atmospheric mass-squared splitting and
acts as the leading contributor to the conversion probability. We can write 
this positive-definite term in the following way
\begin{eqnarray}
\label{eq:Pme_atm}
 &\!\! \!\! \!\! \!\! \!\! \!\! \!\!  P^{\rm {ATM}} &\!\! \simeq\,  4 s_{23}^2 s^2_{13}  \sin^2{\Delta}\,,
 \end{eqnarray}
where $\Delta \equiv  \Delta m^2_{31}L/4E$ is the atmospheric oscillating factor, 
where $E$ is the neutrino energy and $L$ is the baseline. The second and third terms
in Eq.~(\ref{eq:Pme_4nu_3_terms}) are due to interference and can assume
both positive and negative values. The second term is related to the  
solar-atmospheric interference and can be written as
\begin{eqnarray}
 \label{eq:Pme_int_1}
 &\!\! \!\! \!\! \!\! \!\! \!\! \!\! \!\! P^{\rm {INT}}_{\rm I} &\!\!  \simeq\,   8 s_{13} s_{12} c_{12} s_{23} c_{23} (\alpha \Delta)\sin \Delta \cos({\Delta + \delta_{13}})\,.
\end{eqnarray}
The third term comes into the picture as a genuine 4-flavor effect, and is related to the 
atmospheric-sterile interference. It has the following form~\cite{Klop:2014ima} 
\begin{eqnarray}
 \label{eq:Pme_int_2}
 &\!\! \!\! \!\! \!\! \!\! \!\! \!\! \!\! P^{\rm {INT}}_{\rm II} &\!\!  \simeq\,   4 s_{14} s_{24} s_{13} s_{23} \sin\Delta \sin (\Delta + \delta_{13} - \delta_{14})\,.
\end{eqnarray}
From the previous three expressions, we see that the conversion probability depends on three small 
mixing angles: $\theta_{13}$, $\theta_{14}$, and $\theta_{24}$. Interesting to note that the best 
estimates of these three mixing angles (determined in the 3-flavor 
framework~\cite{deSalas:2017kay,Capozzi:2017ipn,Esteban:2016qun} for $\theta_{13}$,
and in the 3+1 scheme~\cite{Capozzi:2016vac,Gariazzo:2017fdh,Dentler:2017tkw} 
for $\theta_{14}$ and $\theta_{24}$) are quite similar and we have 
$s_{13} \sim s_{14} \sim s_{24} \sim 0.15$ (see table~\ref{tab:benchmark-parameters}). 
Therefore, it makes perfect sense to treat these three mixing angles 
$\theta_{13}$, $\theta_{14}$, and $\theta_{24}$ as small quantities having 
the same order $\epsilon$. Conversely, the ratio of the solar over the 
atmospheric mass-squared splitting, 
$\alpha \equiv \Delta m^2_{21}/ \Delta m^2_{31} \simeq \pm \, 0.03$
turns out to be of order $\epsilon^2$. 
From Eqs.~(\ref{eq:Pme_atm})-(\ref{eq:Pme_int_2}), we see that the 
first (leading) term is of the second order in $\epsilon$, 
while both the second and third (subleading) interference terms 
are of the third order in $\epsilon$. Hence, the sizes of the two 
interference terms are expected to be similar.

\section{Details of the numerical analysis}
\label{sec:T2HK}

\subsection{T2HK setup}
\label{sec:T2HKsetup}

The main objective of the long-baseline neutrino program at the proposed 
Hyper-Kamiokande (HK) detector with an intense neutrino beam from the J-PARC 
proton synchrotron is to provide a conclusive evidence for leptonic CP-violation
in neutrino oscillations induced by an irreducible phase $\delta_{13}$ in the 
three-flavor neutrino mixing matrix. This setup is commonly known as ``T2HK''
(Tokai to Hyper-Kamiokande) experiment~\cite{Abe:2011ts,Abe:2014oxa,Abe:2015zbg}. 
To estimate the physics reach of this setup, we closely follow the experimental 
configurations as described in Ref.~\cite{Abe:2014oxa,Abe:2015zbg}. 
The neutrino beam for HK will be produced at J-PARC from the collision of 30 GeV 
protons with a graphite target. In our simulation, we consider an integrated proton 
beam power of 7.5 MW $\times$ $10^7$ seconds, which can deliver in total 
$15.6 \times 10^{21}$ protons on target (p.o.t.) with a 30 GeV proton beam. 
We assume that the T2HK experiment would use 25\% of its full exposure in the
neutrino mode which is $3.9 \times 10^{21}$ p.o.t. and the remaining 75\% 
($11.7 \times 10^{21}$ p.o.t.) would be utilized during antineutrino run.
In this way, we make sure that we have nearly equal statistics in both neutrino 
and antineutrino modes to optimize the search for leptonic CP-violation. 
These neutrinos and antineutrinos will be observed in the gigantic 560 kt (fiducial) 
HK water Cherenkov detector in the Tochibora mine, located 8 km south of 
Super-Kamiokande and 295 km away from J-PARC. The neutrino beamline
from J-PARC is designed to accommodate an off-axis angle of $\sim$ $2.5^\circ$
at the proposed HK site and therefore, the beam peaks sharply at the 
first oscillation maximum of 0.6 GeV to enhance the physics sensitivity.
This off-axis scheme~\cite{Para:2001cu} produces a neutrino beam with 
a narrow energy spectrum which substantially reduces the intrinsic $\nu_e$ 
contamination in the beam and also the background which stems from neutral 
current events. Therefore, the signal-to-background ratio gets improved
significantly. In our analysis, we consider the reconstructed neutrino and 
antineutrino energy range of 0.1 GeV to 1.25 GeV for the appearance channel.
In case of disappearance channel, the assumed energy range is 0.1 GeV to 
7 GeV for both the $\nu_{\mu}$ and $\bar\nu_{\mu}$ candidate events.
We match all the signal and background event numbers following table 19 
and 20 of ref.~\cite{Abe:2014oxa}. The systematic uncertainties play an
important role in estimating the physics sensitivity of the T2HK setup.
Following table 21 of ref.~\cite{Abe:2014oxa}, we consider an uncorrelated 
normalization uncertainty of 3.5\% for both appearance and disappearance
channels in neutrino mode. In case of antineutrino run, the uncorrelated
normalization uncertainties are 6\% and 4.5\% for appearance and
disappearance channels respectively, and also, they do not have any 
correlation with appearance and disappearance channels in neutrino mode.
We assume an uncorrelated 10\% normalization uncertainty on background 
for both appearance and disappearance channels in neutrino and antineutrino 
modes. With all these assumptions on the T2HK setup, we manage to reproduce 
all the sensitivity results which are given in ref.~\cite{Abe:2014oxa}. 
Here, we would like to mention that according to the latest report by the 
Hyper-Kamiokande Proto-Collaboration~\cite{Abe:2016ero}, the total beam
exposure is $27 \times 10^{21}$ p.o.t. and the fiducial mass for the proposed 
HK detector is 374 kt. Comparing the exposures in terms of (kt $\times$ p.o.t.), 
we see that our exposure is 1.156 times smaller than the exposure that has been
considered in ref.~\cite{Abe:2016ero}. But, certainly, the results presented in this 
work would not change much and the conclusions drawn based on these results
would remain valid even if we consider the new exposure as reported in 
ref.~\cite{Abe:2016ero}.

\subsection{Statistical Method}
\label{sec:simulation-details}

\begin{table}[t]
\begin{center}
{
\newcommand{\mc}[3]{\multicolumn{#1}{#2}{#3}}
\newcommand{\mr}[3]{\multirow{#1}{#2}{#3}}
\begin{tabular}{|c|c|c|}
\hline\hline
\mr{2}{*}{\bf Parameter} & \mr{2}{*}{\bf True Value} & \mr{2}{*}{\bf Marginalization Range} \\
  & &  \\
\hline\hline
\mr{2}{*}{$\sin^2{\theta_{12}}$} & \mr{2}{*}{0.304} & \mr{2}{*}{Not marginalized} \\
  & &  \\
\hline
\mr{2}{*}{$\sin^22\theta_{13}$} & \mr{2}{*}{$0.085$} & \mr{2}{*}{Not marginalized} \\ 
  & &  \\
\hline
\mr{2}{*}{$\sin^2{\theta_{23}}$} & \mr{2}{*}{0.50} & \mr{2}{*}{[0.34, 0.68]} \\
  & &  \\
\hline
\mr{2}{*}{$\sin^2{\theta_{14}}$} & \mr{2}{*}{0.025} & \mr{2}{*}{Not marginalized} \\
  & &  \\  
\hline
\mr{2}{*}{$\sin^2{\theta_{24}}$} & \mr{2}{*}{0.025} & \mr{2}{*}{Not marginalized} \\
  & &  \\ 
\hline
\mr{2}{*}{$\sin^2{\theta_{34}}$} & \mr{2}{*}{0.0} & \mr{2}{*}{Not marginalized} \\
  & &  \\  
\hline
\mr{2}{*}{$\delta_{13}/^{\circ}$} & \mr{2}{*}{[- 180, 180]} & \mr{2}{*}{[- 180, 180]} \\
  & &  \\
\hline
\mr{2}{*}{$\delta_{14}/^{\circ}$} & \mr{2}{*}{[- 180, 180]} & \mr{2}{*}{[- 180, 180]} \\
  & &  \\
\hline
\mr{2}{*}{$\delta_{34}/^{\circ}$} & \mr{2}{*}{0} & \mr{2}{*}{Not marginalized} \\
  & &  \\  
\hline
\mr{2}{*}{$\frac{\Delta{m^2_{21}}}{10^{-5} \, \rm{eV}^2}$} & \mr{2}{*}{7.50} & \mr{2}{*}{Not marginalized} \\
  & &  \\
\hline
\mr{2}{*}{$\frac{\Delta{m^2_{31}}}{10^{-3} \, \rm{eV}^2}$ (NH)} & \mr{2}{*}{2.475} &\mr{2}{*}{Not marginalized} \\
  & & \\
\hline
\mr{2}{*}{$\frac{\Delta{m^2_{31}}}{10^{-3} \, \rm{eV}^2}$ (IH)} & \mr{2}{*}{- 2.4} &\mr{2}{*}{Not marginalized} \\
  & & \\
\hline
\mr{2}{*}{$\frac{\Delta{m^2_{41}}}{\rm{eV}^2}$} & \mr{2}{*}{1.0} & \mr{2}{*}{Not marginalized} \\
  & &  \\
\hline\hline
\end{tabular}
}
\caption{Oscillation parameters considered in our analysis. The second column
depicts the true values of the oscillation parameters used to simulate the ``observed'' 
data set. The third column shows the ranges over which $\sin^2\theta_{23}$, $\delta_{13}$,
and $\delta_{14}$ are varied while minimizing the $\chi^{2}$ to produce the final results.}
\label{tab:benchmark-parameters}
\end{center}
\end{table}

In this section, we briefly describe the numerical technique and analysis 
procedure that we adopt to produce our sensitivity results. To compute
the experimental sensitivities, we rely on the well-known GLoBES 
software~\cite{Huber:2004ka,Huber:2007ji} along with its new physics tools 
which estimates the median sensitivity of the experiment without including
statistical fluctuations. We make appropriate changes in the 
$\nu_{\mu} \to \nu_{e}$ and $\nu_{\mu} \to \nu_{\mu}$ transition probabilities 
to incorporate the 3+1 scheme with one light eV-scale sterile neutrino. 
We do the same for antineutrino as well. We observe that for the small 
values of mixing angles $\theta_{14}$ and $\theta_{24}$ considered in the 
present work (see table~\ref{tab:benchmark-parameters}), the 4-flavor 
$\nu_{\mu} \to \nu_{\mu}$ disappearance probability is very similar to the 
3$\nu$ case, which agrees with the findings in ref.~\cite{Klop:2014ima}.
In this paper, the strategy that we adopt for the statistical treatment, is exactly 
similar to what have been discussed in section 4 of ref.~\cite{Agarwalla:2016mrc}. 
Table~\ref{tab:benchmark-parameters} depicts the true values of the oscillation 
parameters and their marginalization ranges which we consider in our simulation. 
Our benchmark choices for the three-flavor neutrino oscillations parameters are 
in close agreement with the values obtained in the recent global fit 
studies~\cite{deSalas:2017kay,Capozzi:2017ipn,Esteban:2016qun}.
As far as the atmospheric mixing angle is concerned, we consider
maximal mixing\footnote{Latest 3$\nu$ global 
fits~\cite{deSalas:2017kay,Capozzi:2017ipn,Esteban:2016qun} prefer 
non-maximal $\theta_{23}$ with two nearly degenerate solutions:
one $<$ $45^{\circ}$, known as lower octant (LO), and the 
other $>$ $45^{\circ}$, termed as higher octant (HO). But, maximal
mixing is still allowed in 3$\sigma$ range.} ($45^{\circ}$) as the 
true choice, and in the fit, we marginalize over the range given in 
table~\ref{tab:benchmark-parameters}. We keep $\theta_{13}$ 
fixed in the fit because we have already achieved very good 
precision on this parameter, and the Daya Bay experiment 
is going to improve the precision on $\theta_{13}$ further 
by the end of its run~\cite{Ling:2016wgq}.
The proposed T2HK experiment is also very sensitive to this 
parameter and will be able to restrict $\theta_{13}$ in a narrow 
range in the fit, minimizing its impact in marginalization. 

As far as $\Delta m_{31}^2$ is concerned, we have 
already achieved a very good precision ($\sim$ 2\%)
on this parameter. Also, T2HK will accumulate huge 
amount of disappearance events (governed by 
$\nu_\mu \to \nu_\mu$ oscillation channel), and 
therefore, it will be highly sensitive to the value of 
$\Delta m_{31}^2$, and will be able to measure 
it with very high precision. Moreover, the proposed
JUNO experiment has the potential to significantly 
improve our measurements of $\theta_{12}$, 
$\Delta m_{21}^2$, and $\Delta m_{31}^2$ to a precision 
of less than 1\%~\cite{An:2015jdp}. For all these reasons, 
we keep the value of $\Delta m_{31}^2$ fixed in the fit. 

We present all the sensitivity results assuming normal 
hierarchy\footnote{We have checked that 
the results do not differ much if we generate the prospective data 
with inverted hierarchy (IH).} (NH) as the true choice. For the 
CP-violation searches and the reconstruction of the CP phases,
we show our results with and without marginalizing over both the 
choices of hierarchy in the fit, which enable us to see how much 
the sensitivities can be deteriorated due to the possible degeneracies 
between sgn($\Delta m^2_{31}$) and CP phases. While showing the 
results for the octant measurement of $\theta_{23}$, we marginalize 
over both the choices of hierarchy in the fit. This issue of hierarchy 
marginalization in the fit becomes irrelevant in case of the mass 
hierarchy discovery studies where our aim is to exclude the wrong
hierarchy in the fit. The true value of $\delta_{13}$ is varied in its
allowed range of $[-\pi, \pi]$, and it has been marginalized over its 
full range in the fit if the analysis demands so. Following the Preliminary 
Reference Earth Model (PREM)~\cite{PREM:1981}, we take the 
line-averaged constant Earth matter density of 2.8 g/cm$^{3}$ for the 
T2HK baseline. 

In our simulation, we consider the new mass-squared splitting 
$\Delta m^2_{41}$ to be 1 eV$^2$ as preferred by the recent 
short-baseline data. Here, we would like to mention that our 
sensitivity results would remain the same for other values of this
parameter, provided that $\Delta m^2_{41} \gtrsim 0.1\,$eV$^2$.
The rapid oscillations induced by such large values of 
$\Delta m^2_{41}$ get completely averaged because of the finite
resolution of the detector. Due to the same reason, the T2HK 
setup is insensitive to the sign of $\Delta m^2_{41}$ and we can
safely assume it to be positive. As far as the active-sterile mixing
angles are concerned, we take the true values of 0.025 for both
$\sin^2\theta_{14}$ and $\sin^2\theta_{24}$.
Our choices for these new mixing angles are quite close
to the best-fit values obtained by the global 
3+1 fits~\cite{Capozzi:2016vac,Gariazzo:2017fdh,Dentler:2017tkw}. 
Needless to mention that one requires a huge amount
of computational resources to marginalize over these 
new mixing angles in their present allowed ranges in the fit 
along with all the CP phases. In light of the limited computational 
resources that we have at present, we decide to keep the values 
of these new mixing angles fixed at their true values in the fit, 
and we devote our computational power more towards the 
CP phases, which long-baseline experiments can only probe.
We vary the true value of $\delta_{14}$ in its allowed range of 
[$-\pi$, $\pi$] and it has been marginalized over its full range 
in the fit as required. We consider $\theta_{34}$ = 0 and 
$\delta_{34}$ = 0 in our simulations%
\footnote{In vacuum, $\nu_{\mu} \to \nu_e$ oscillation 
probability is independent of $\theta_{34}$ (and $\delta_{34}$).
For the T2HK baseline, the matter effect is very small and in the 
probability expression we see a very tiny dependence on these 
parameters which can be safely neglected. An analytical 
understanding of this issue is given in the appendix of 
ref.~\cite{Klop:2014ima}.}.

In our analysis, we do not explicitly simulate the near detector
of T2HK which may provide some information regarding 
$\theta_{14}$ and $\theta_{24}$, but surely, the near detector
data is blind to the CP phases which we want to explore in this
work. While producing our sensitivity results, we perform a full 
spectral analysis using the binned events spectra for the 
T2HK setup. Following refs.~\cite{Huber:2002mx,Fogli:2002pt},
we use the well-known ``pull'' method to marginalize the 
Poissonian $\Delta\chi^{2}$ over the uncorrelated systematic
uncertainties. We quote the statistical significance\footnote{For 
a detailed discussion on the statistical interpretation of oscillation 
experiments, see the recent 
refs.~\cite{Ciuffoli:2013rza,Blennow:2013oma,Blennow:2013kga,Elevant:2015ska}.} 
of our results for 1 d.o.f. in terms of $n\sigma$, where $n=\sqrt{\Delta\chi^2}$.

 \begin{figure}[t!]
\centerline{
 \includegraphics[height=8.cm,width=7.5cm]{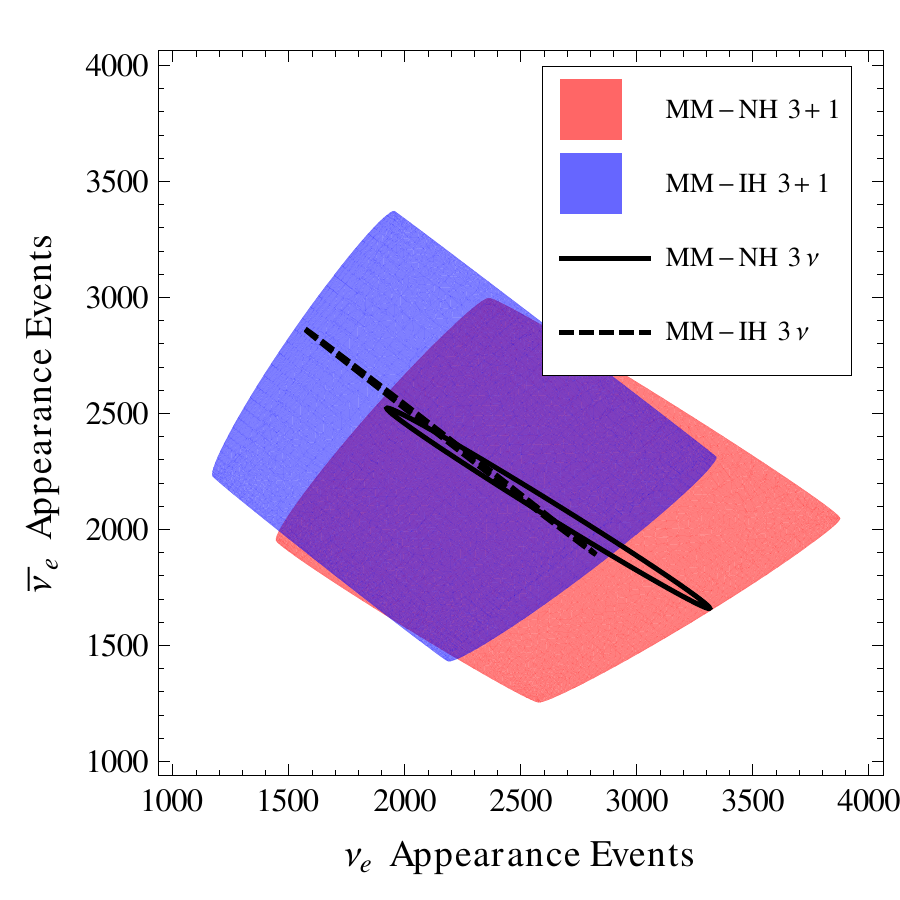}
 \includegraphics[height=8.cm,width=7.5cm]{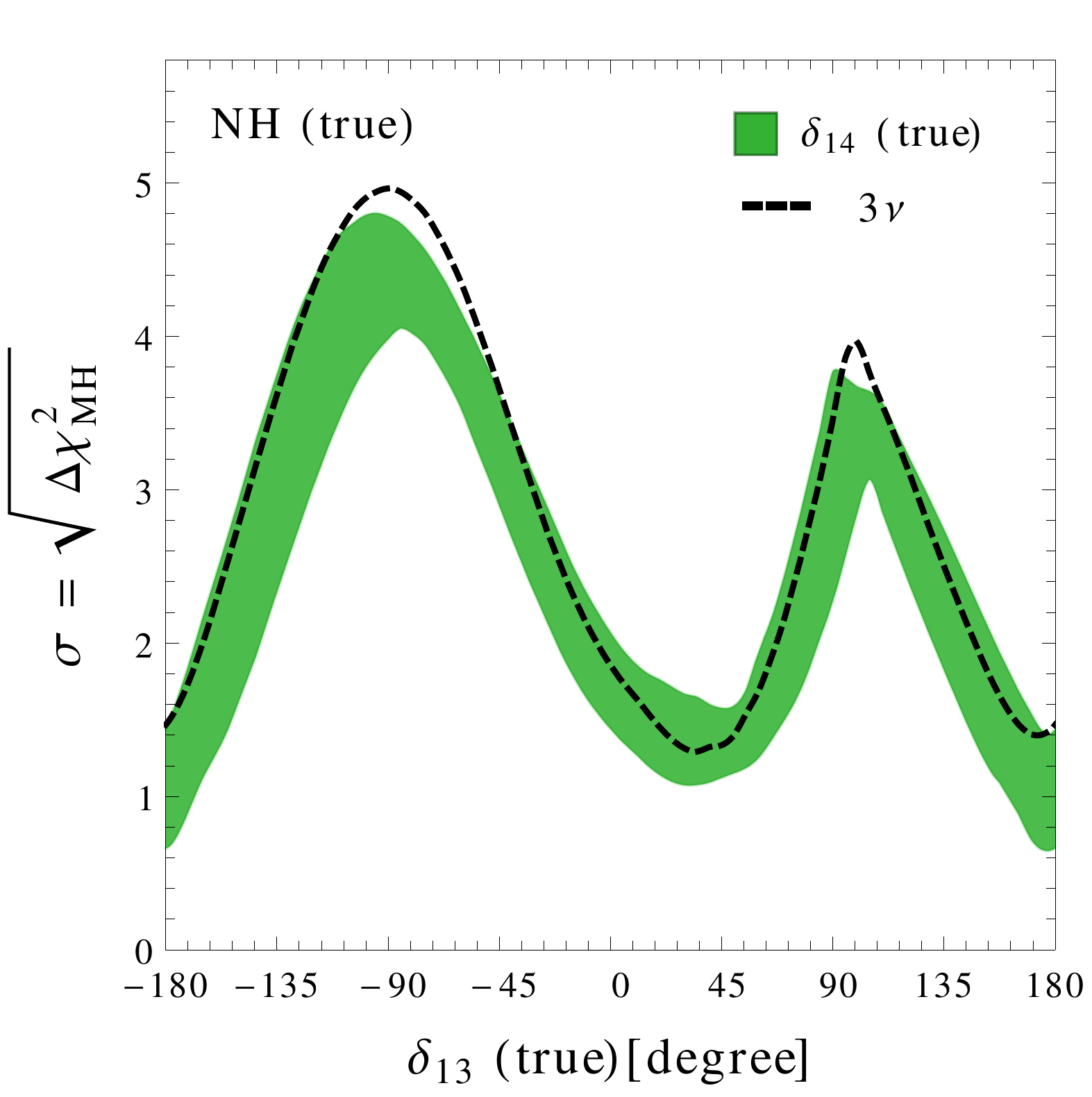} 
 }
  \caption{Left panel: bievents plot for 3-flavor (black ellipses) and 4-flavor case (colored blobs).
 Right panel: Discovery potential for identifying  the correct hierarchy (NH) as a function 
of true $\delta_{13}$. We have fixed true and test $\theta_{14} = \theta_{24} = 9^0$.
The black dashed line represents the 3-flavor case, while the green band corresponds
 to the 3+1 scheme. Such a band is obtained by varying the true value of $\delta_{14}$ in 
 the range $[-\pi,\pi]$ and marginalizing the test value of $\delta_{14}$ in the same range.}
 \label{fig:mh} 
 \end{figure}

\section{Mass hierarchy discovery potential in the 3+1 scheme}
\label{sec:MH}

In this section, we treat the sensitivity of T2HK to the neutrino mass hierarchy. 
First of all we provide a discussion at the level of the electron neutrino events 
making use of  the bievents plots. Then, we present the results of the full analysis. 

The left panel of Fig.~\ref{fig:mh}, represents the bi-event plots
where the two axes report the number of $\nu_e$ ($x$-axis)
 and  $\bar \nu_e$ ($y$-axis) events.
The two ellipses correspond to the 3-flavor case and are obtained varying 
the CP phase $\delta_{13}$ in the range $[-\pi,\pi]$. The solid (dashed) ellipse
refers to the NH (IH). The offset between the two ellipses is a direct consequence
of matter effects, which act in opposite directions in the transition probability
for the two cases of NH and IH. This offset confers to the experiments T2HK 
some sensitivity to the mass hierarchy. However, differently from DUNE 
(where matter effects are stronger), the separation of the two ellipses
is not complete and one expects that the MH discovery potential is 
limited. In the 3+1 scheme, there are two CP phases and their variation in the
range $[-\pi,\pi]$ gives much more freedom. The scatter plots obtained 
varying  the CP phases $\delta_{13}$ and $\delta_{14}$ are superimposed
to the 3-flavor ellipses in the left panel Fig.~\ref{fig:mh}.
Despite the larger parameter freedom, we can observe that the 
separation of two blobs corresponding to NH and IH remains similar 
to that of the 3-flavor ellipses. Therefore, we expect that the discovery
potential of the MH in the 3+1 scheme will  suffer only a mild deterioration
with respect to the 3-flavor case.

 \begin{figure}[t!]
\hspace{0.2cm}
\centerline{
 \includegraphics[height=10.cm,width=16.5cm]{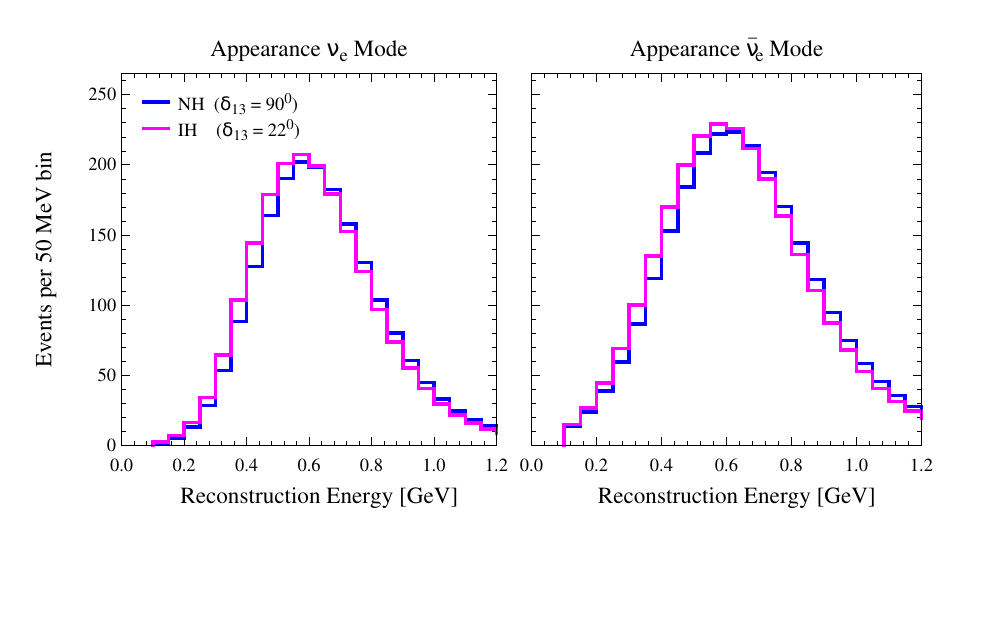}}
 \vspace{-2cm}
  \caption{Energy spectra for neutrinos (left panel) and antineutrinos (right panel)
 for NH (blue lines) and IH (magenta lines). The values of the CP-phase for NH and IH 
 have been chosen so that the total number of neutrino and antineutrino events are basically 
 identical for the two hierarchies. In this case, the noticeable residual sensitivity to MH 
 ($4 \sigma$ according to right panel of Fig.~1 for $\delta_{13} = 90^0$) is entirely
 provided by the energy shape information.}
 \label{fig:mh_spectrum} 
 \end{figure}
 
In the right panel of Fig.~\ref {fig:mh}, we plot the MH discovery potential as a function of the true value of $\delta_{13}$. 
We define the discovery potential as the statistical significance at which the wrong
test hierarchy is rejected given a data set which is generated with the true hierarchy.
Let us consider first the 3-flavor scenario. In this case the sensitivity to the MH (for NH 
as the true hierarchy choice) is represented by a black dashed line. Such a curve is obtained
by marginalizing the test value of $\delta_{13}$ in the range $[-\pi,\pi]$ and 
$\theta_{23}$ (test) within the $3\sigma$ range allowed by the global
analyses~\cite{deSalas:2017kay,Capozzi:2017ipn,Esteban:2016qun}. Although
such a 3-flavor MH sensitivity curve has been already shown in several previous 
works~\cite{Abe:2016ero,Liao:2016orc,Fukasawa:2016yue,Ballett:2016daj,Ghosh:2017ged,Raut:2017dbh,Agarwalla:2017nld,Choubey:2017cba,Chakraborty:2017ccm,C.:2014ika}, its behavior
deserves some clarification. In fact, on the basis of the bivents plot on the left panel one
would expect maximal sensitivity for $\delta_{13} = -90^0$, since the distance of 
the representative point on the (solid) NH ellipse (right lower point) has maximal distance from the IH ellipse.
While this elementary feature is confirmed by the right panel Fig.~\ref {fig:mh}, which presents 
a pronounced maximum ($\sim 5\sigma$) for $\delta_{13} \simeq -90^0$, the same figure also shows something unexpected.
Quite surprisingly a good sensitivity  ($\sim 4\sigma$) appears also for  $\delta_{13} \simeq 90^0$.
However, for such a value of $\delta_{13}$, the bievents plot (see the ellipses in the left panel) 
shows that the distance of the representative point on the solid NH ellipse (left upper point) 
from the (dashed) IH ellipse is very close to zero.
This is a well known fact (known as $\delta_{13}$-MH degeneracy) which is supposed
to reduce to zero the sensitivity to the MH. So the question arises where does the $4\sigma$
sensitivity found in the right panel Fig.~\ref {fig:mh}  for $\delta_{13} = 90^0$ come from. The answer 
is that although the number of events cannot distinguish NH from IH for $\delta_{13} = 90^0$, 
the full energy spectrum can.
In Fig.~\ref{fig:mh_spectrum} we elucidate this fact by showing, for the 3-flavor case, 
the neutrino (left panel) and antineutrino (right panel) spectra for the two choices 
$(\mathrm{NH},\, \delta_{13} = 90^0)$ and $(\mathrm{IH},\, \delta_{13} = 22^0)$, which correspond approximately to
a common point in the 3-flavor bievents plot (i.e. equal number of neutrino and antineutrino events). 
Figure~\ref{fig:mh_spectrum} clearly shows that the two spectra are different and therefore the spectral
information provides sensitivity to the MH discrimination. 
Therefore, we conclude that the spectral information has a crucial role in 
the sensitivity to the MH and is able to break the $\delta_{13}$-MH degeneracy
for a wide range of the values of $\delta_{13}$ around $90^0$.
Unfortunately, for values of $\delta_{13}$ around $45^0$ and $180^0$,
the spectral information is less effcetive and the sensitivity does not surpass the
$2\sigma$ level. As far as we know, the role of the spectral information in the MH
discrimination in T2HK has not been discussed before in the literature.
 
In the 3+1 scheme (green band in the right panel of Fig.~\ref {fig:mh}), in addition to $\delta_{13}$ and $\theta_{23}$,
the marginalization is carried out over the CP phase $\delta_{14}$. 
We have taken $\theta_{14} = \theta_{24} = 9^0$ both in data and fit.
The spread of the band is due to the variation of the true value of $\delta_{14}$ in its entire range of $[-\pi,\pi]$.
We observe that the qualitative behavior of the 3+1 band follows that of the 3-flavor case, and
as expected, there is only a mild deterioration of the MH discovery potential. 
Also in this case the spectral information plays an important role in the MH discrimination.
We can conclude that although the MH discovery potential of T2HK is quite limited%
\footnote{Although T2HK has a limited sensitivity to the mass hierarchy, Hyper-Kamiokande (HK)
can settle this issue using the atmospheric neutrinos at more than 3$\sigma$ C.L. for both NH and IH
provided that $\sin^2 \theta_{23}>0.45$~\cite{Yokoyama:2017mnt}. Combining beam and atmospheric neutrinos in 
HK, the mass hierarchy can be determined at more than $3\sigma$ (5$\sigma$) with five (ten) 
years of data~\cite{Yokoyama:2017mnt}.}
it is rather robust with respect to the perturbations induced by a light sterile neutrino species.%
 
\section{CP-violation searches in the 3+1 scheme}
\label{sec:CPV}
   
In this section, we analyze the capability of T2HK of nailing down the enlarged 
CPV sector implied by the 3+1 scheme. Before showing the full numerical results we  
present a discussion at the level of bievents plots, which will serve as a valid guide
to understand the results of the full simulations.  As a first step we assess the sensitivity
to the CPV induced by the CP phase $\delta_{13}$ in the 3-flavor framework.
As a second step we show how the sensitivity to the CPV induced by $\delta_{13}$ changes 
in the 3+1 scheme. 
As a third step, we investigate the sensitivity to the non-standard CP phase $\delta_{14}$ (which
appears only in the 3+1 scheme).
Throughout the discussion, for the first time in the literature, we evidence the essential role of 
the energy spectrum in providing a minimal guaranteed sensitivity to the CPV induced by
the new phase $\delta_{14}$. Finally, we discuss the capability of reconstructing the true values 
of the two phases $\delta_{13}$ and $\delta_{14}$.

 \begin{figure}[t!]
\centerline{
 \includegraphics[height=5. cm,width=5.cm]{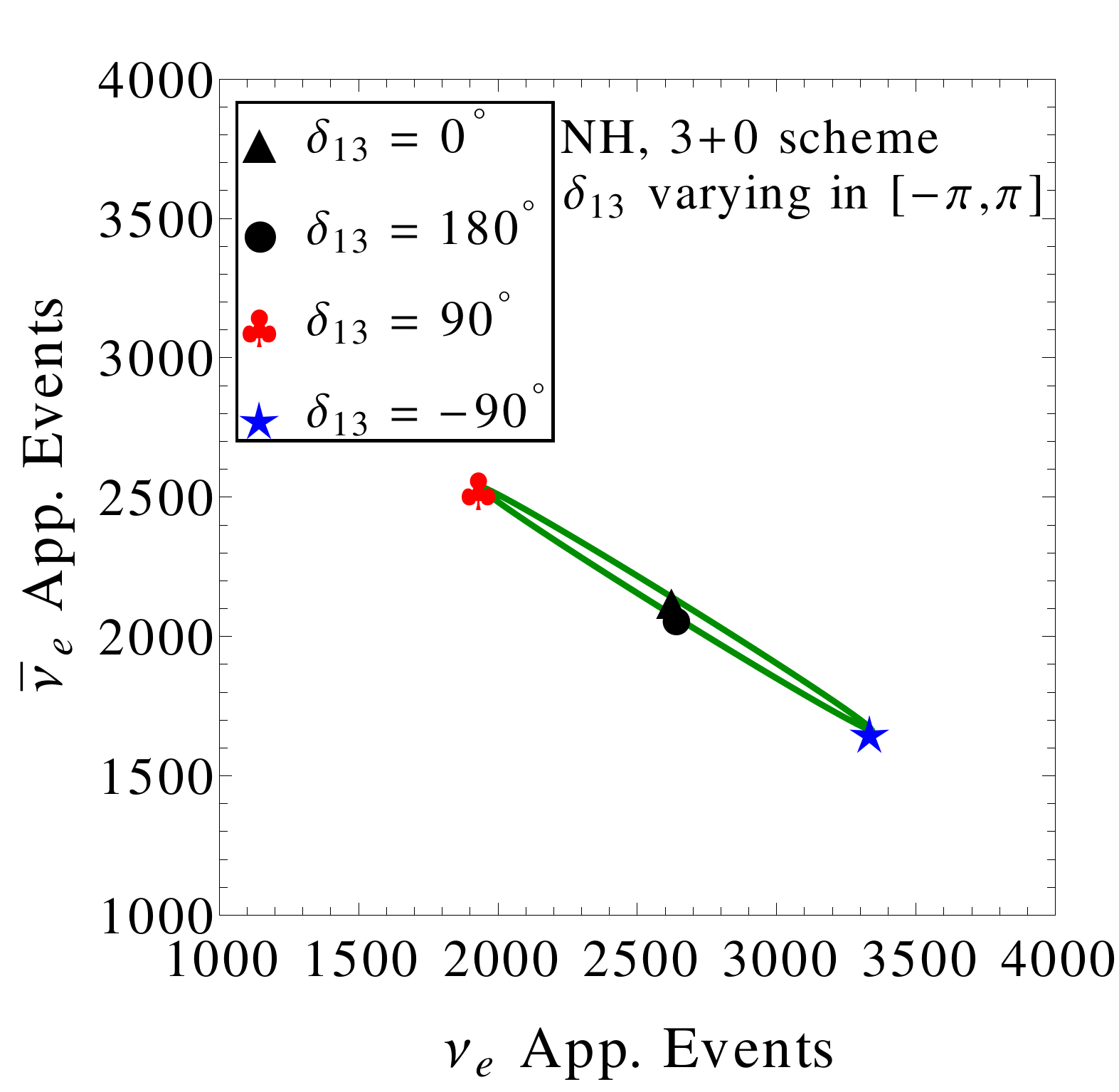}
  \includegraphics[height=5. cm,width=5.cm]{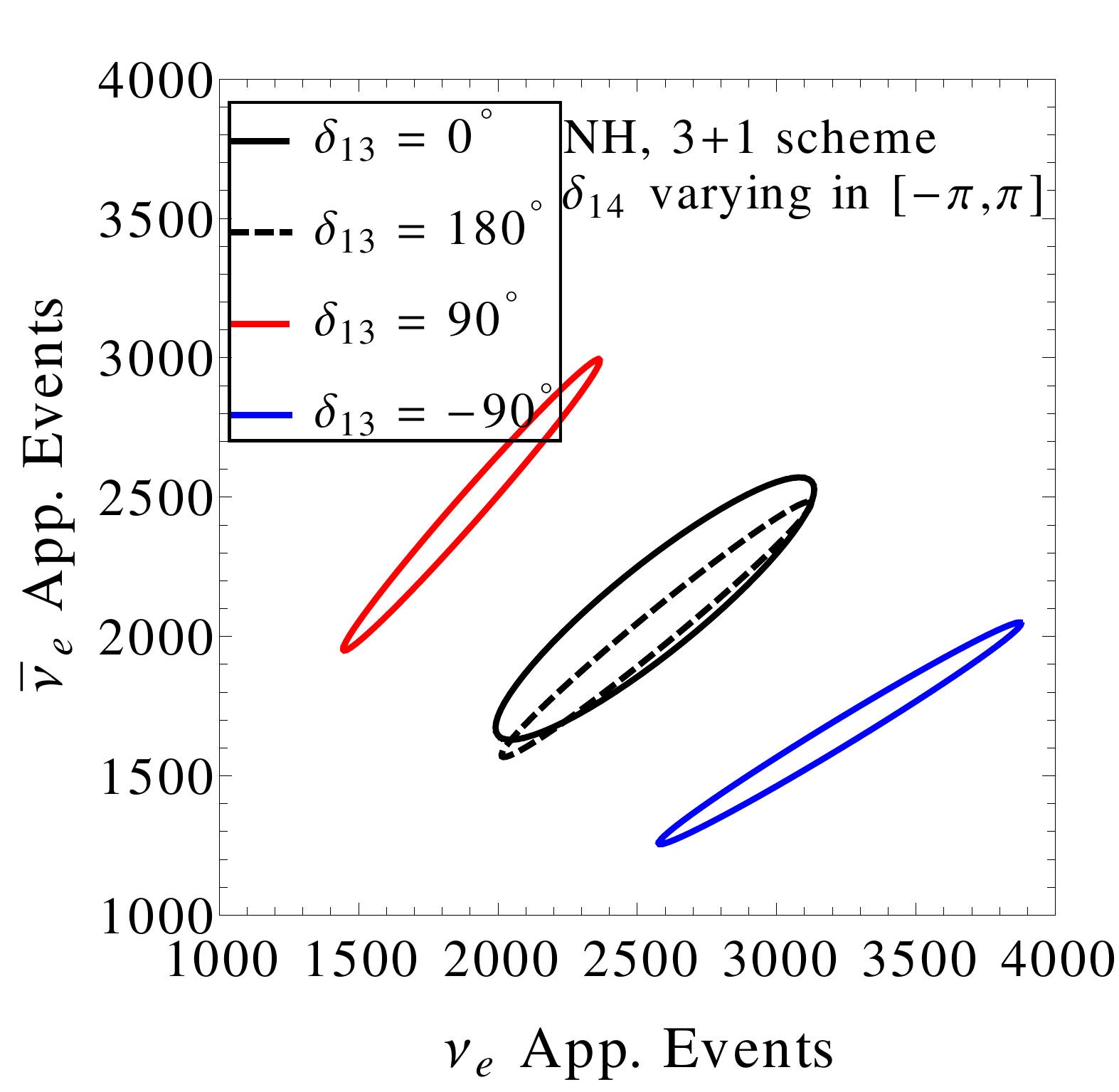}
 \includegraphics[height=5. cm,width=5.cm]{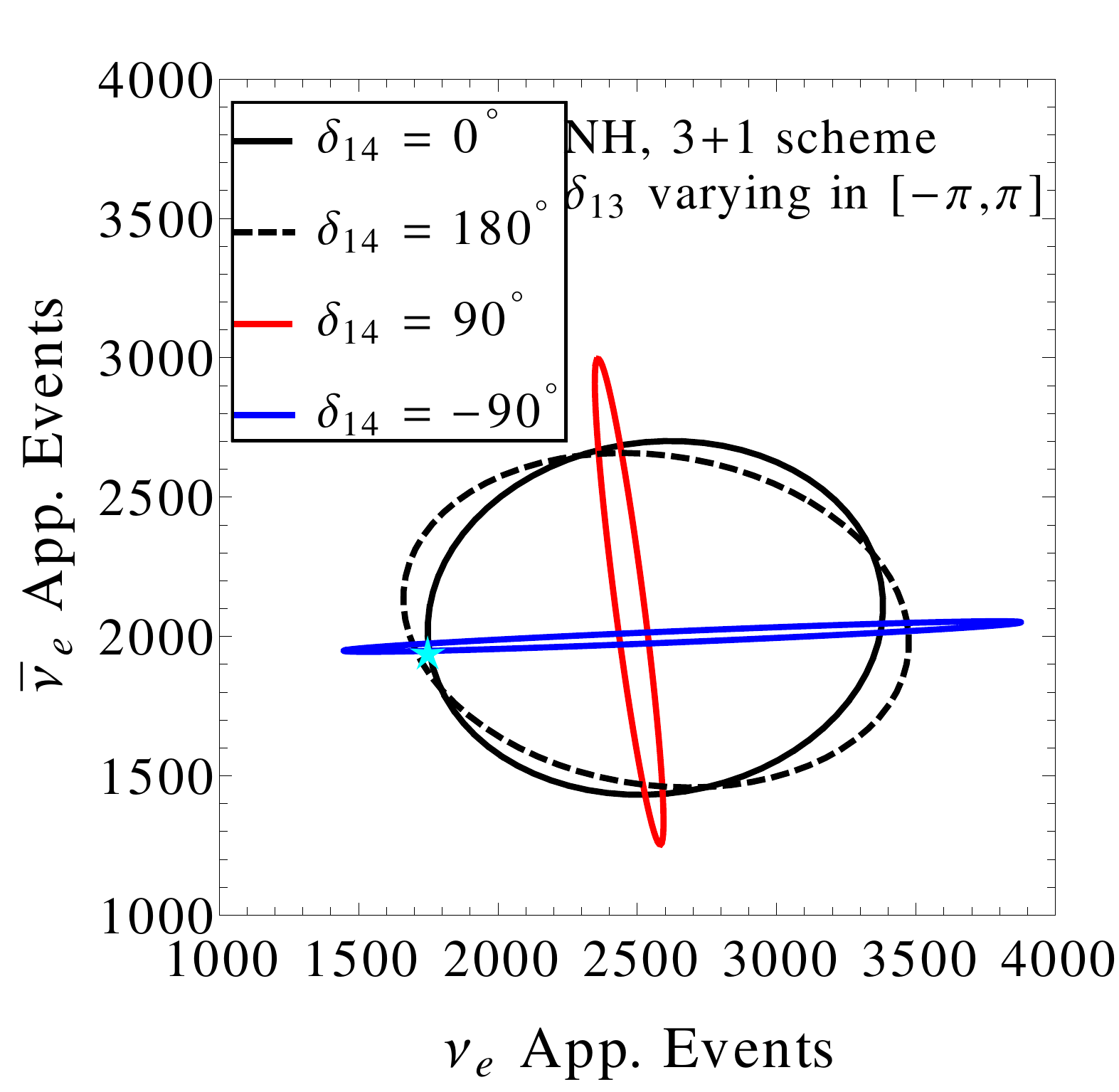}
  }
 \caption{Bievents plots for the 3-flavor framework (left panel) and 3+1 scheme (central and right panels).
  In the left panel two black marks are located for the test cases of no CPV ($\delta_{13} = 0,\pi$) 
  and two colored ones for the true cases corresponding to maximal CPV  ($\delta_{13} = -\pi/2,\pi/2$).   
  For completeness we draw also the (green) ellipse where the 3-flavor model lives, which is obtained 
  by varying $\delta_{13}$ in the range $[-\pi,\pi]$. The non-zero distance between the black marks
  and the colored ones indicates that events counting can determine CPV induced by the phase $\delta_{13}$
  in the 3$\nu$ case. In the 3+1 scheme a fixed value of $\delta_{13}$ is represented by an ellipse,
  where $\delta_{14}$ varies in the range $[-\pi,\pi]$.
  The non-zero minimal distance between the black ellipses and each of the two colored ones indicates
  that events counting is sensitive to CPV induced by the phase $\delta_{13}$ also in the 3+1 case. 
  The right panel illustrates the sensitivity to the CPV induced by $\delta_{14}$. In this case 
  four ellipses are plotted for four fixed values of $\delta_{14}$ (while $\delta_{13}$ is varying in the range $[-\pi,\pi]$).
   Each of the two ellipses (blue and red) 
  corresponding to maximal CPV induced by $\delta_{14}$ crosses 
  the two ellipses (solid and dashed black) corresponding to no CPV induced by $\delta_{14}$. 
   In the (unlucky) crossing points the events counting is not sensitive to CPV induced by the new CP-phase $\delta_{14}$.
   See the text for details. }
 \label{CPV_bievents}
 \end{figure}

 \begin{figure}[t!]
\centerline{
 \includegraphics[height=7.5 cm,width=7.5cm]{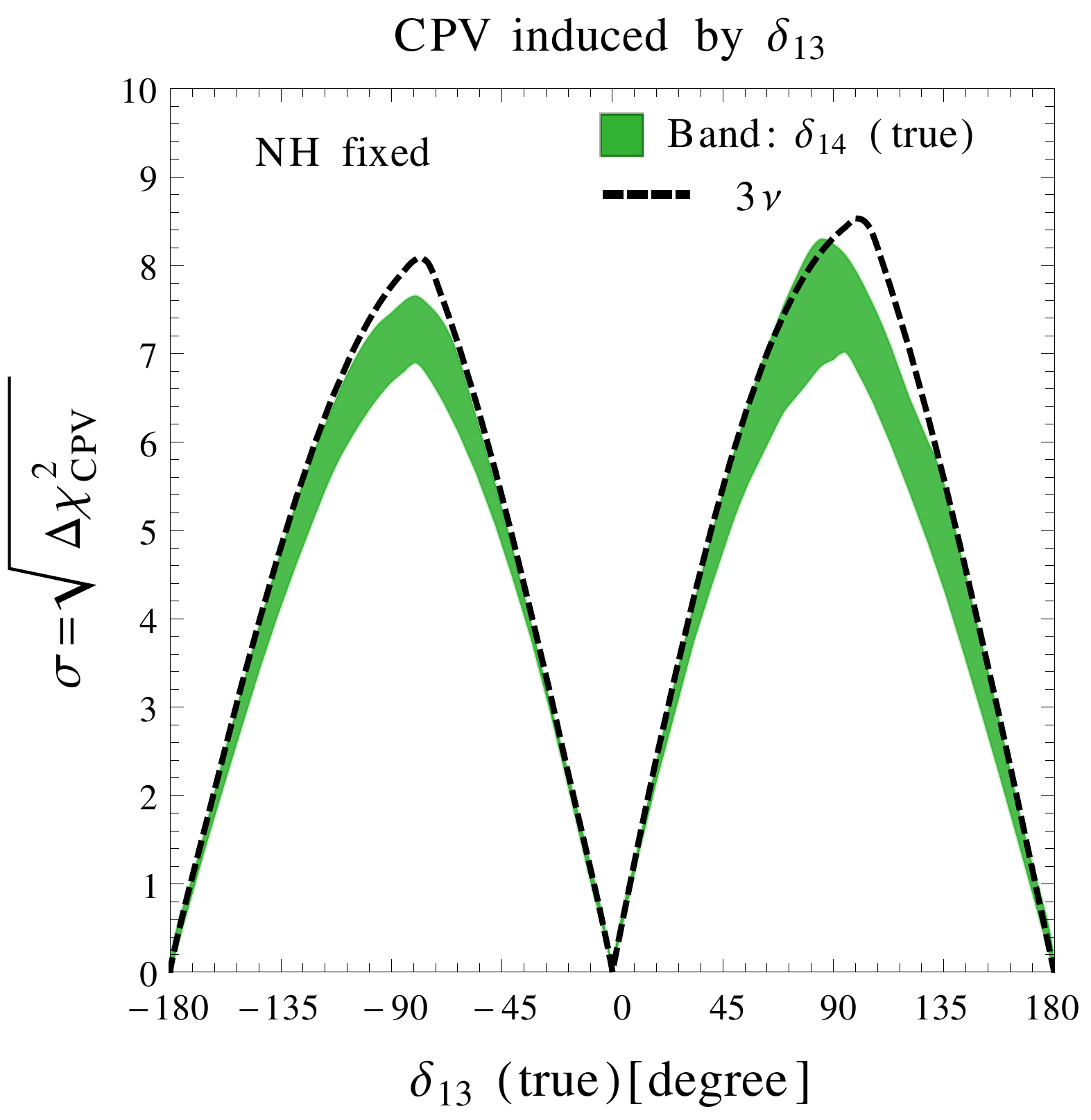}
 \includegraphics[height=7.5 cm,width=7.5cm]{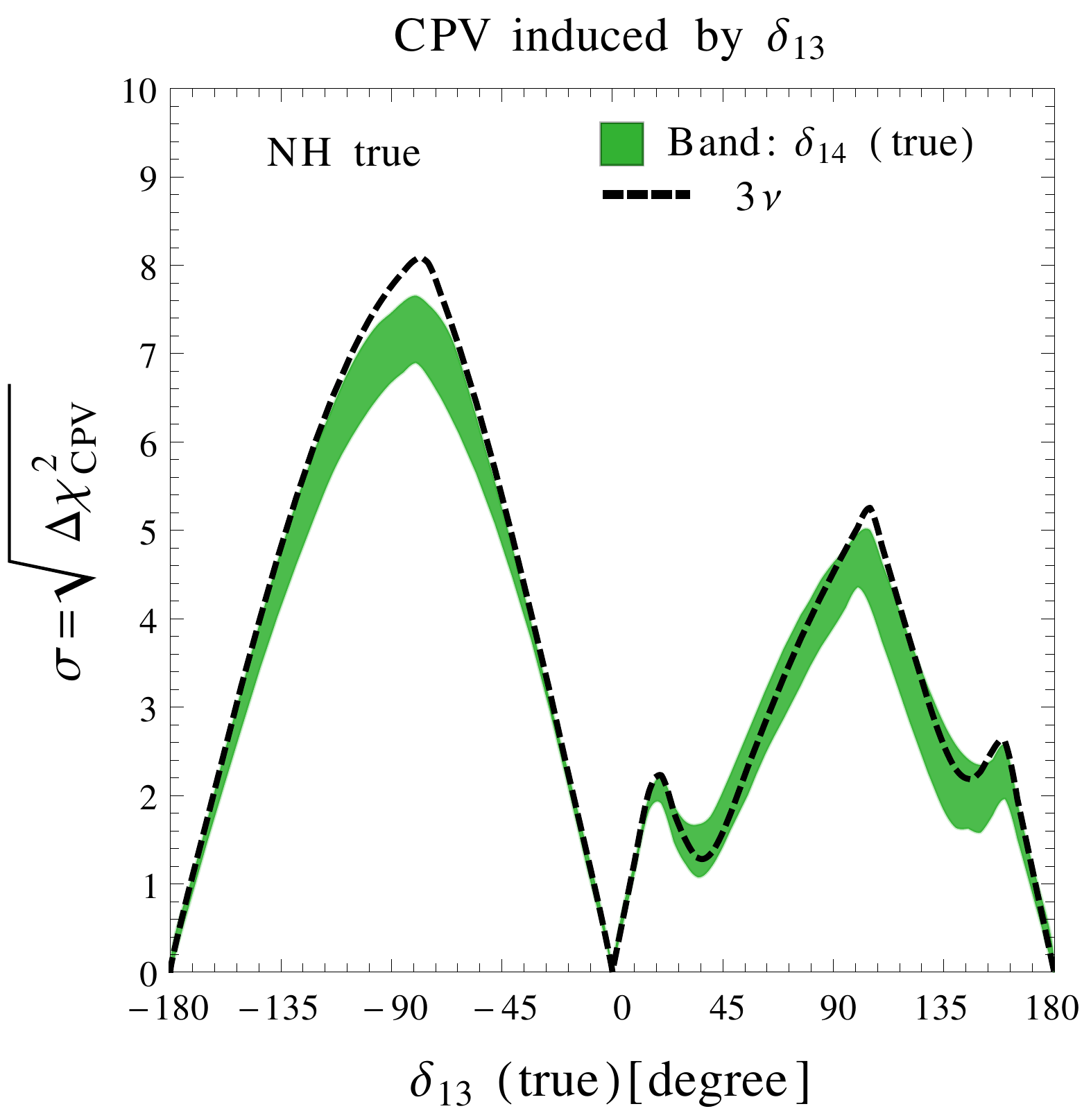}
  }
\caption{T2HK discovery potential of  $\delta_{13} \ne (0,\pi)$. In the left panel 
  the MH is fixed to be the NH (both true and test value). In the right panel the MH is marginalized with NH true. 
   In both panels, the black dashed curve corresponds to the 3-flavor case. 
In 3+1 scenario, we fix the true and test values of $\theta_{14}$ = $\theta_{24}$ = $9^0$.
In both panels,  the colored bands are obtained in the 3+1 scheme by varying the unknown 
value of the true $\delta_{14}$ in its entire range of $[-\pi,\pi]$ while marginalizing over test 
$\delta_{14}$ in the same range.}
  \label{CPV_delta_13}
 \end{figure}

\subsection{CP-violation discovery potential}

The sensitivity of CPV induced by a given (true) value of a CP phase $\delta_{ij}^{\mathrm{true}}$ 
is defined as the statistical significance at which one can exclude the test hypothesis of no CPV, i.e.
the (test) cases $\delta_{ij}^{\mathrm{test}}=0$ and $\delta_{ij}^{\mathrm{test}}=\pi$. 
The qualitative behavior of the sensitivity to the CPV induced by a given phase $\delta_{ij}$ can
be traced by a careful inspection of opportune bievents plots.
The three panels of Fig.~\ref{CPV_bievents} illustrate graphically the following three
cases: i) CPV induced by $\delta_{13}$ in the 3-flavor framework (left panel);
ii) CPV induced by $\delta_{13}$ in the 3+1 scheme (central panel);
iii) CPV induced by $\delta_{14}$ in the 3+1 scheme (right panel).
In all panels we have assumed NH for definiteness. In the standard 3-flavor scenario
(left panel) there is only one (true) CP-phase ($\delta_{13}^{\mathrm{true}}$), which is the running parameter
over the green ellipse in the range $[-\pi,\pi]$. The two (inequivalent) CP conserving
test points ($\delta_{13}^{\mathrm{test}} = 0,\,\pi$) are indicated with a triangle and a circle.
We also indicate two points of the ellipse corresponding to maximal CP violation
($\delta_{13}^{\mathrm{true}} = -\pi/2,\,\pi/2$) with a star and a black club. According to the definition
of the sensitivity given at the beginning of this subsection, the non-zero distance between 
any of the two (maximally) CP violating (true) points from both the CP conserving (test) points, guarantees that, 
in the 3-flavor scheme, events counting is sensitive to CPV induced by $\delta_{13}$.
 The central panel of Fig.~\ref{CPV_bievents} concerns the 3+1 scheme and is meant
to illustrate the sensitivity to CPV induced by $\delta_{13}$ in such an enlarged scheme.
We recall that in the 3+1 scheme a given value of $\delta_{13}$ corresponds to an ellipse,
whose running parameter is the other phase $\delta_{14}$ in its range $[-\pi,\pi]$
(for a detailed study of the properties of the 3+1 ellipses see~\cite{Agarwalla:2016mrc}).
 The central panel reports four 
ellipses of which two (the black ones) correspond to the (inequivalent) CP conserving test cases 
($\delta_{13}^{\mathrm{test}} = 0,\,\pi$) and the other two (the colored ones) represent maximal CP 
violation ($\delta_{13}^{\mathrm{true}} = -\pi/2,\,\pi/2$). In the 3+1 scheme,
in order to decide if there is sensitivity to CPV (induced by $\delta_{13}$) one has
to look at the minimal distance between a generic  point lying on one of the two colored ellipses
(this point will correspond to a generic value of $\delta_{14}^{\mathrm{true}}$) from the
two black ellipses (where $\delta_{14}^{\mathrm{test}}$  runs in the range $[-\pi,\pi]$).
From the plot, we see that the major axes of the four ellipses are almost parallel, and
as a consequence whatever is the point chosen on the colored ellipses (i.e., the value
of  $\delta_{14}^{\mathrm{true}}$), its distance from the black ellipses is always non-zero. 
Also, me may observe that the minimal distance between colored and black ellipses is very similar
(just slightly lower because the ellipses are not exactly parallel) to that found in the 3-flavor scheme
 between the CP violating cases (star, black club) and the CP conserving ones (triangle and circle).
Therefore, we expect that in the 3+1 scheme, the sensitivity to CPV induced by $\delta_{13}$ 
will be only slightly lower than that achieved in the standard 3-flavor scenario.
Finally, the right panel of  Fig.~\ref{CPV_bievents}  illustrates the sensitivity to the CPV induced by $\delta_{14}$. In this case 
the four ellipses correspond to four fixed values of $\delta_{14}$ (while $\delta_{13}$ is varying in the range $[-\pi,\pi]$).
Differently form the central panel, now the four ellipses have a completely different behavior. 
In particular, the two test CP conserving black ellipses $(\delta_{14}^{\mathrm{test}} =0,\, \pi)$
are almost circular, while the two colored ellipses corresponding to maximal CP violation 
$(\delta_{14}^{\mathrm{true}} =-\pi/2,\, \pi/2)$ are almost degenerate
with a line and are orthogonal to each other (this behavior was already found and discussed
in~\cite{Agarwalla:2016mrc}). As a result, the distance of a generic point located on one of the two colored
ellipses (which will have a generic value of  $\delta_{13}^{\mathrm{true}}$)  from the two black ellipses is not necessarily 
bigger than zero. In fact, each colored ellipse crosses a black ellipse in four points. Each of the crossing 
points between a colored ellipse and a black one will correspond to a particular pair of
values of the two phases ($\delta_{13}^{\mathrm{true}}$,  $\delta_{13}^{\mathrm{test}}$).
In such points the numbers of neutrino and antineutrino events are identical for
maximal CPV and no CPV induced by $\delta_{14}$ and, as a consequence, events counting
is completely insensitive to the CPV induced by $\delta_{14}$.
We have checked that such ``unlucky pairs" approximately correspond to the sixteen possible 
combinations in couples of the four values $\delta_{13} = (-135^0, -45^0, 45^0, 135^0)$.%
\footnote{Note that in twelve of the sixteen combinations $\delta_{13}^{\mathrm{true}}$
is much different from $\delta_{13}^{\mathrm{test}}$. Hence a prior determination
of $\delta_{13}$ with a good precision, imposing $\delta_{13}^{\mathrm{test}} \simeq \delta_{13}^{\mathrm{true}}$,
would reduce the crossing points only two four.} 

In Fig.~\ref{CPV_delta_13}, we report the discovery potential of CPV induced by $\delta_{13}$. 
In the left panel we have assumed that the hierarchy is 
known a priori and is NH. In the right panel we assume no a priori knowledge of the hierarchy 
generating the data with NH as the true choice and marginalizing over the two hierarchies.   
In this last case, the task of the experiment is more difficult because it has to identify both CPV and the MH. 
In both the panels, the black dashed curves correspond to the 3-flavor scenario
and the green band to the 3+1 scheme. In the 3+1 scenario, we fix the true and test 
values of $\theta_{14}$ and $\theta_{24}$ to be $9^0$. In both the panels, the green
bands are obtained by varying the unknown true value of $\delta_{14}$
in the range of $[-\pi,\pi]$ while marginalizing over its test value in the same range. 
In the left panel two
maxima are present around $\delta_{13}$ $\sim$ $\pm$ $90^0$.  In the right panel 
the height of the second maximum around $\delta_{13}$ $\sim$ $90^0$
is drastically reduced as a result of the degeneracy between MH and the CP phase $\delta_{13}$.
Finally, we observe that the deterioration found in the 3+1 scheme is mild
in agreement with the discussion made at the bievents level.

  \begin{figure}[t!]
\centerline{
 \includegraphics[height=7.5 cm,width=7.5cm]{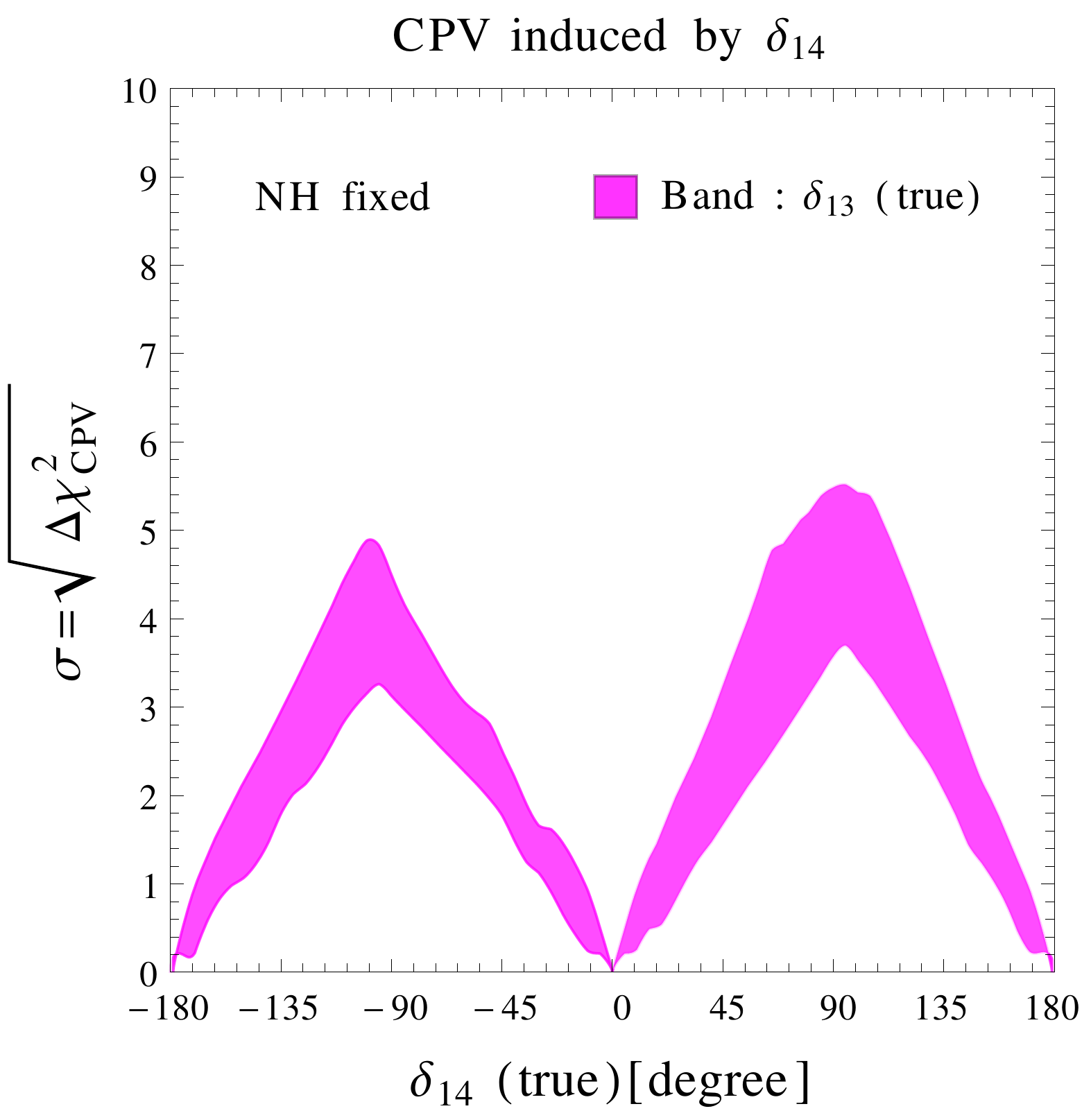}
 \includegraphics[height=7.5 cm,width=7.5cm]{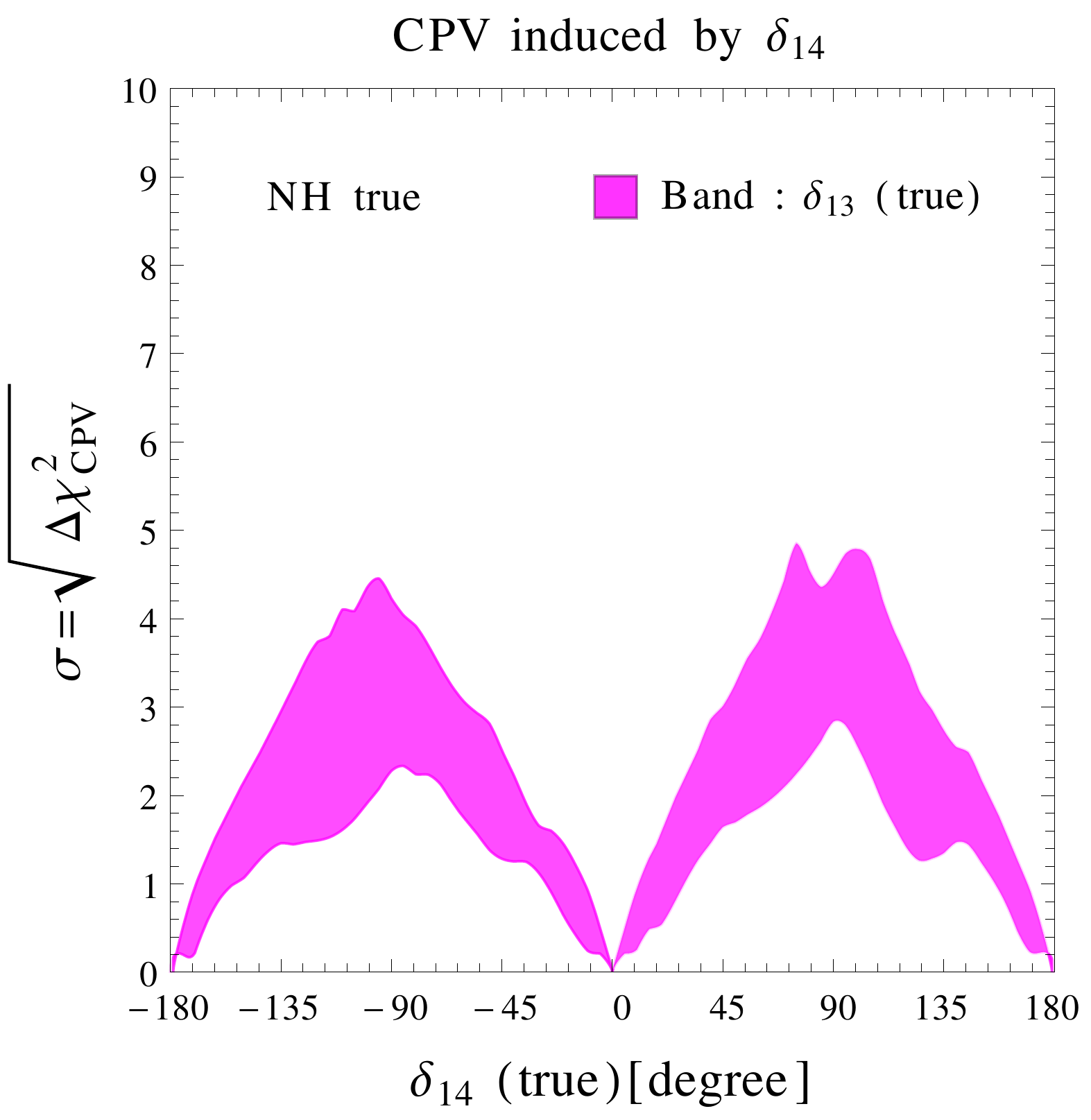}
 }
  \caption{Discovery potential of  $\delta_{14} \ne (0,\pi)$.
 In the left panel the MH is fixed to be the NH. In the right panel the MH is 
 marginalized with NH as the true choice. 
 In both cases we have taken true and test $\theta_{14} = \theta_{24} = 9^0$. 
 The bands have been obtained by varying the true value of
$\delta_{13}$ in its allowed range of $[-\pi,\pi]$ 
while marginalizing over its test value in the same range.
}
 \label{CPVvsdel14}
 \end{figure}
 
 \begin{figure}[t!]
\hspace{0.2cm}
\centerline{
 \includegraphics[height=10.cm,width=16.5cm]{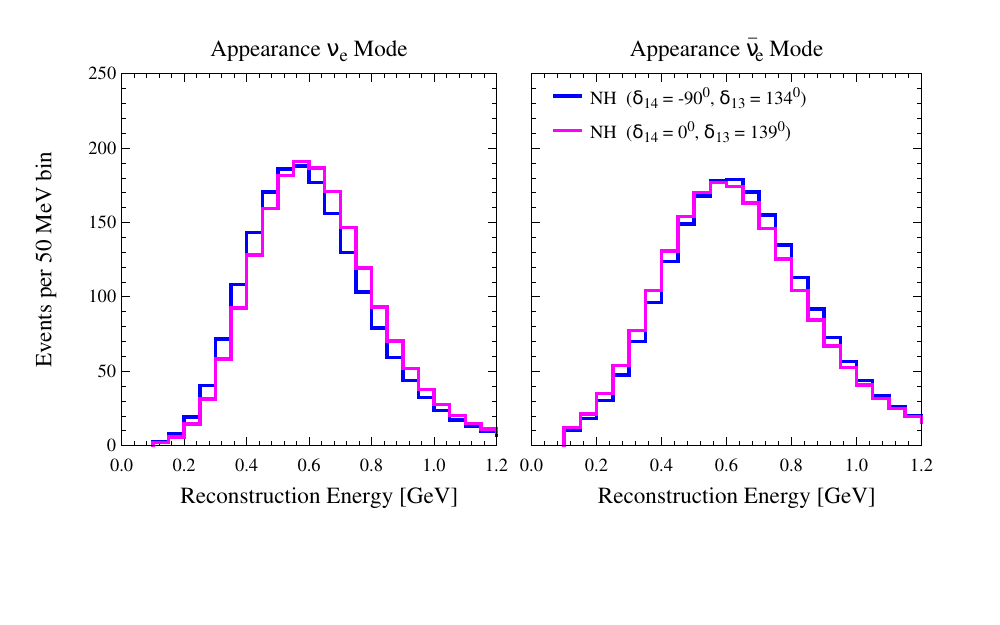}}
 \vspace{-2cm}
  \caption{Energy spectra for neutrinos (left panel) and antineutrinos (right panel)
 obtained for NH. The magenta histograms correspond to a case of no CPV induced by $\delta_{14}$ ($\delta_{14} =0$).
 The blue histograms correspond to a case of maximal CPV induced by $\delta_{14}$ ($\delta_{14}=-90^0$).
 The two values of $\delta_{13}$ have been chosen in order to have complete degeneracy
 at the level of both neutrino and antineutrino events. These choices of phases correspond to
 one of the sixteen crossing points in the right panel of Fig.~\ref{CPV_bievents} (therein indicated by a cyan star).
 The difference in the spectra is in this case the only source of sensitivity to the CPV
 induced by $\delta_{14}$, which is  $\sim 3.5 \sigma$ according to the left panel of Fig.~5 for $\delta_{14} = -90^0$.}
 \label{fig:CPV_delta_14_spectrum} 
 \end{figure}
 
In Fig.~\ref{CPVvsdel14}, we display the discovery potential of CPV induced 
by $\delta_{14}$. Like in the previous figure, in the left panel the hierarchy is
fixed (NH) while in the right panel it is unknown (assuming the true hierarchy to be normal).
In each panel, the bands have been obtained by varying the true values of
the CP phase $\delta_{13}$ in the range $[-\pi,\pi]$  while marginalizing over their
test values in the same range in the fit. Based on the discussion made at the
bievents level one should expect that for unlucky values of
($\delta_{13}^\mathrm{true},\, \delta_{13}^\mathrm{test}$) the sensitivity 
drops to zero. In contrast, the left panel of Fig.~\ref{CPVvsdel14} shows that there is a
guaranteed minimal sensitivity, which for the maximally CPV cases $\delta_{14}^\mathrm{true} = \pm 90^0$
is roughly $3.5\sigma$. This means that even in the unlucky
cases of complete degeneracy at the level of the total (neutrino and antineutrino) events 
between the case of maximal CPV and that of no CPV, a residual sensitivity is provided by the energy spectral information.
In Fig.~\ref{fig:CPV_delta_14_spectrum} we elucidate this fact by showing
the neutrino (left panel) and antineutrino (right panel) spectra for two 
cases that are degenerate at the events level.  The first one corresponds to maximal CPV
induced by $\delta_{14}$ ($\delta_{14} = -90^0,\, \delta_{13} = 134^0)$ and the second one to no CPV
induced by the same phase $(\delta_{14} = 0,\, \delta_{13} =139^0)$. This choice of the phases
correspond to one of the sixteen crossing points in the bievents plot in the right panel of Fig.~\ref{CPV_bievents} 
(therein indicated by a cyan star). 
Figure~\ref{fig:CPV_delta_14_spectrum} clearly shows that, although the number of 
(both neutrino and antineutrino) events is identical,
 the two spectra are different and therefore they provide some sensitivity to the CPV induced by $\delta_{14}$. 
As far as we know, the role of the spectral information in the CPV
induced by the new CP phase $\delta_{14}$ has not been discussed before in the literature.
Finally, we note that,  there is a deterioration of the discovery
potential to CPV induced by $\delta_{14}$ when going from the left panel 
(known hierarchy) to the right panel (unknown hierarchy) of Fig.~\ref{CPVvsdel14}.
Also in this case,
the spectral information guarantees a minimal degree of sensitivity,
which for the maximally CPV cases $\delta_{14}^\mathrm{true} = \pm 90^0$
is roughly $2.5\sigma$.

\begin{figure}[t!]
\hspace{0.5cm}
\centerline{
 \includegraphics[width=13.1cm]{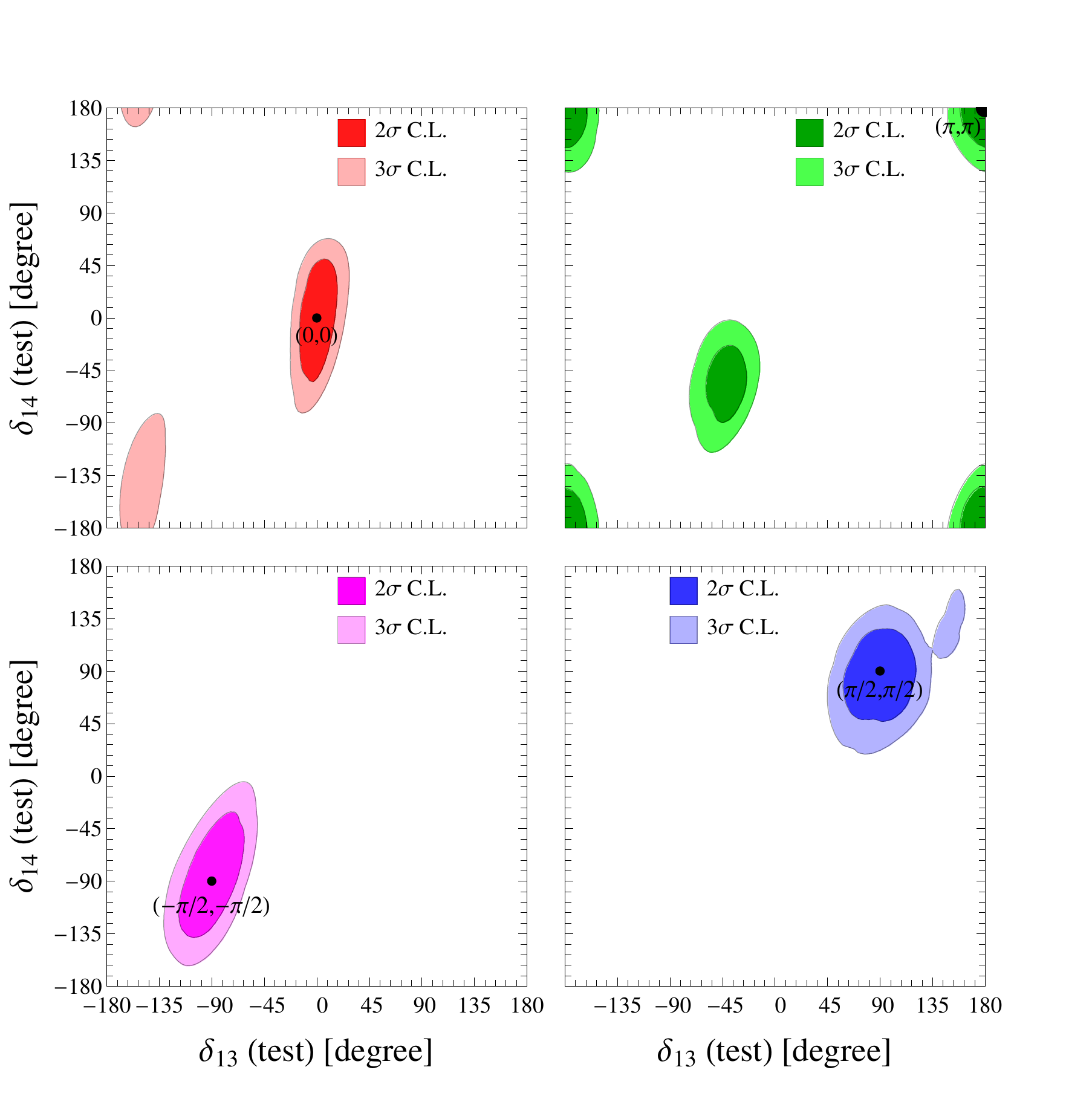}
 }
 \vspace{-0.5cm}
  \caption{Reconstructed regions for the two CP phases $\delta_{13}$ and $\delta_{14}$ 
for the four benchmark pairs of their true values indicated in each panel. We have taken
the NH as the true hierarchy and we have marginalized over the two possible hierarchies
in the test model. The contours refer to 2$\sigma$ and 3$\sigma$ levels.}
 \label{CPV_rec_1}
 \end{figure}
 
\begin{figure}[t!]
\hspace{0.0cm}
\centerline{
 \includegraphics[width=13.0cm]{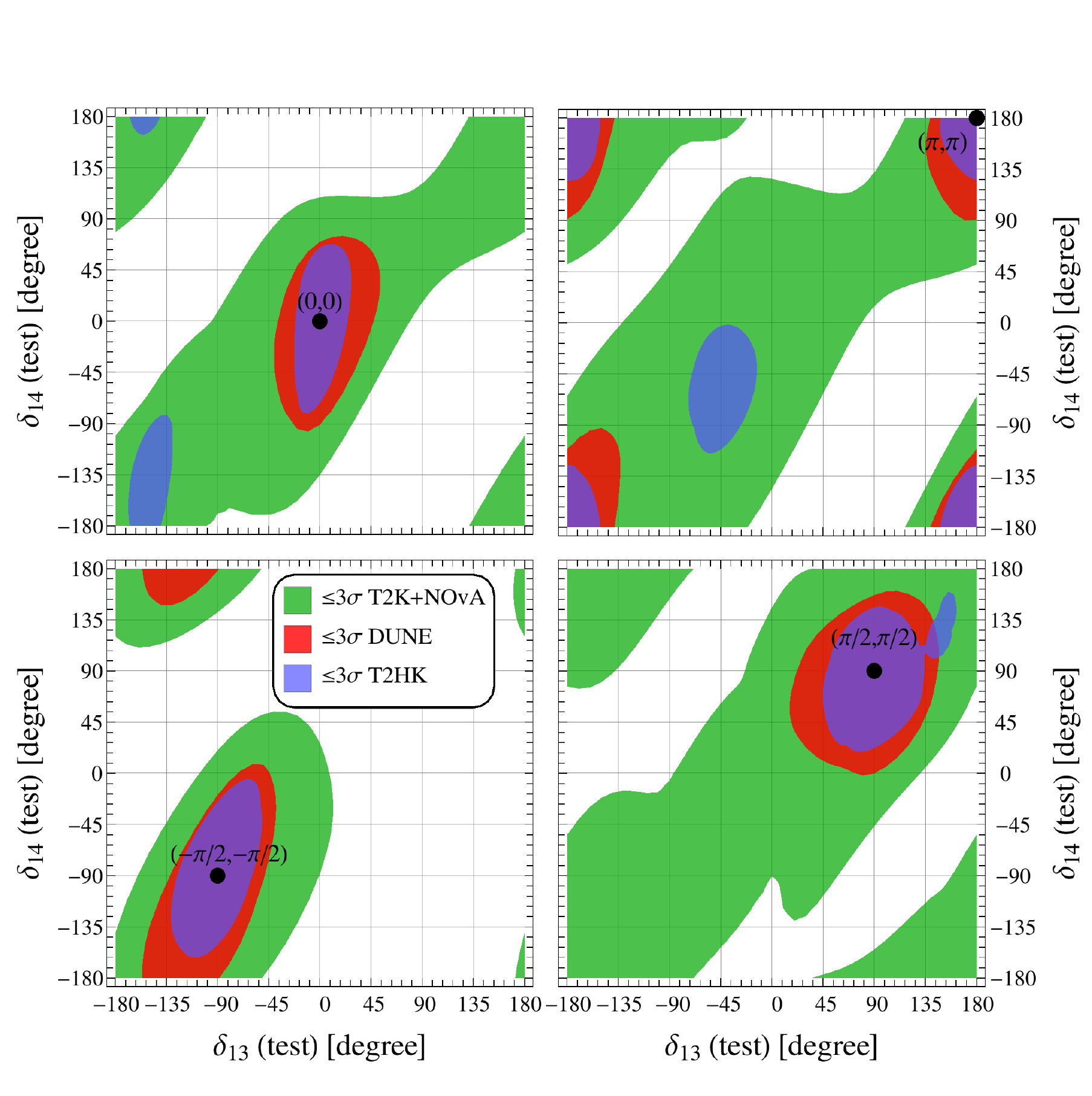}
 }
  \caption{Reconstructed regions for the two CP phases $\delta_{13}$ and $\delta_{14}$ 
for three different experimental setups. We have taken
the NH as the true hierarchy and we have marginalized over the two possible hierarchies
in the test model. The contours correspond to the 3$\sigma$ level.}
 \label{CPV_rec_2}
 \end{figure}

\subsection{Reconstruction of the CP phases}

Until now, we have focused on the discovery potential of the CPV 
generated by the two CP phases $\delta_{13}$ and $\delta_{14}$.
Here, we investigate the capability of reconstructing the true values
of the two CP phases. With this purpose we consider the four benchmark
cases displayed in Fig.~\ref{CPV_rec_1}. The two upper panels represent the 
CP-conserving scenarios $[0,0]$ and $[\pi,\pi]$. The third and fourth panels represent 
two CP-violating cases $[-\pi/2, -\pi/2]$ and $[\pi/2, \pi/2]$. 
In each panel, we draw the regions reconstructed around  the true values 
of $\delta_{13}$ and $\delta_{14}$. In this plot we have taken the NH as the 
true hierarchy and we have marginalized over the two possible hierarchies in the test model.
The two confidence levels correspond to 2$\sigma$ and 3$\sigma$ (1 d.o.f.).
The typical 1$\sigma$ level uncertainty on the reconstructed CP phases 
is approximately $15^0$ ($30^0$) for $\delta_{13}$ ($\delta_{14}$). 
We see however that spurious islands appear in three of the four cases considered.
In first panel an island appears around $(\delta_{13},\delta_{14}) = (-150^0, -150^0)$,
in the second panel around%
\footnote{We recall that both $\delta_{13}$ and $\delta_{14}$ are cyclic variables,
therefore the four corners in the second panel of  Fig.~\ref{CPV_rec_1} form
a unique region.}
 $(\delta_{13},\delta_{14}) = (-45^0, -45^0)$, 
in the fourth panel around $(\delta_{13},\delta_{14}) = (150^0, 150^0)$.
This misreconstruction is imputable to the well known degeneracy between 
$\delta_{13}$ and sgn$(\Delta m^2_{31})$. The combination of phases chosen
for the third panel seems to be the most favorable one (no misreconstructed 
islands). This happens because in such a case the difference in the number of events
for NH and IH is more pronounced and therefore there is a better discrimination
of the MH (see the discussion in~\cite{Agarwalla:2016mrc}). We have explicitly checked
that if the MH is supposed to be known a priori (i.e. it is fixed and not marginalized in the fit),
the spurious islands disappears in all cases.
Therefore, our results show that in T2HK in order to have good reconstruction 
capabilities of the two CP phases, one needs the prior knowledge of 
the mass hierarchy.

We close this section by performing a comparison of the CP-phase 
reconstruction potential of three different experimental setups%
\footnote{For a detailed discussion of the CP-phases reconstruction potential 
of T2K+NO$\nu$A and DUNE, see respectively~\cite{Agarwalla:2016mrc} and~\cite{Agarwalla:2016xxa}.} 
: T2K+NO$\nu$A, DUNE and T2HK. 
In Fig.~\ref{CPV_rec_2} we plot the reconstructed regions obtained by these three 
setups for the same benchmark values of the true phases chosen in Fig.~\ref{CPV_rec_1}.
For visual clearness we report only the 3$\sigma$ contours.
The plots clearly show that there is huge gain when going from T2K+NO$\nu$A
to the new generation experiments DUNE and T2HK. Concerning these two last
experiments, the figure shows that T2HK is slightly more precise than DUNE in reconstructing
the region around the true values CP phases. However, while in T2HK there are 
small spurious islands, this does not happen in DUNE. This different behavior is
rooted in the fact that the degeneracy between $\delta_{13}$ and 
sgn$(\Delta m^2_{31})$ is less pronounced in DUNE. This, in turn, happens
because DUNE can neatly separate the two hierarchies thanks to the
larger matter effects.

\section{Sensitivity to the octant of $\theta_{23}$}
\label{sec:octant}

\begin{figure}[t!]
\centerline{
 \includegraphics[height=9. cm,width=17.cm]{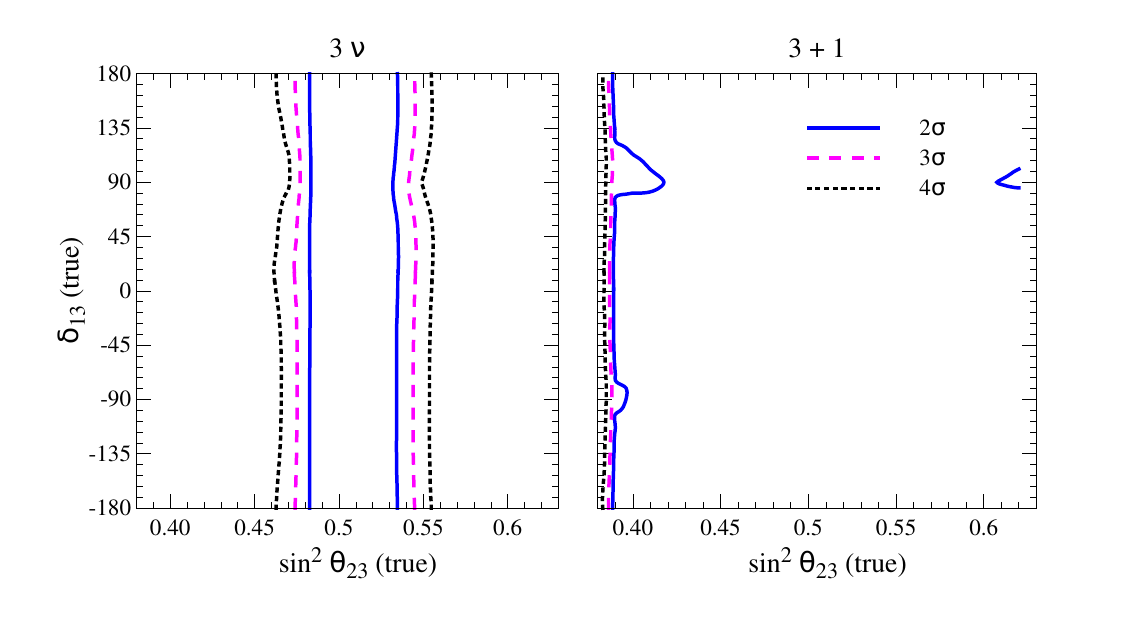}
}
  \caption{Discovery potential for identifying the correct octant assuming NH as true choice.
The left (right) panel corresponds to the 3$\nu$ (3+1) case.
In the 3$\nu$ case, we marginalize away ($\theta_{23}, \delta_{13}$) (test). In the 3+1 case, 
in addition, we marginalize away also $\delta_{14}$ (true) and $\delta_{14}$ (test) fixing
$\theta_{14} = \theta_{24} = 9^0$. }
 \label{fig:octant}
 \end{figure}

The latest global fits of neutrino data indicate a preference for non-maximal $\theta_{23}$ with two nearly degenerate
solutions, one in the lower octant ($\theta_{23} <\pi/4$), and the other in the higher octant ($\theta_{23} >\pi/4$).
The resolution of this octant ambiguity is a crucial target of next-generation LBL experiments.
In a recent work~\cite{Agarwalla:2016xlg} it was shown that in the 3+1 scheme the sensitivity of the future LBL experiment DUNE 
to the the $\theta_{23}$ octant can be deteriorated in a drastic way. 
Here we perform a similar analysis to check if the same conclusion holds for T2HK.
Figure~\ref{fig:octant} displays the discovery potential for identifying the true octant in the plane [$\delta_{13}, \sin^2 \theta_{23}$] (true) 
assuming NH as true choice. The left (right) panel represents the results obtained in  3$\nu$ (3+1) scheme. 
In the 3+1 case we marginalized over the CP phase $\delta_{14}$ (true) 
(in addition to all the test parameters) since such a phase is unknown. Hence, the outcome of this
procedure provides the minimal guaranteed sensitivity. The three contours correspond, 
respectively, to 2$\sigma$, 3$\sigma$ and 4$\sigma$ confidence levels (1 d.o.f.). The comparison 
of the two panels neatly shows that in the 3+1 scheme no minimal sensitivity is guaranteed in the entire plane. 
We have checked that similar results are valid also in the case of IH as true MH. Hence, we confirm that
also in T2HK, like in DUNE, the identification of the octant of $\theta_{23}$ is problematic when one
works in the enlarged 3+1 framework.  

\section{Conclusions and Outlook}
\label{Conclusions}

We have investigated the capability of T2HK to provide information on the mass hierarchy,
the  CP phases, and the octant of $\theta_{23}$, in the presence of one light sterile neutrino species.
Working in 3+1 scheme, we have found that the discovery potential of the mass hierarchy
is rather robust with respect to the perturbations induced by the sterile species. 
We have also found that the discovery potential of CPV induced
by the standard CP phase $\delta_{13}$ gets only slightly deteriorated compared to the standard 3$\nu$ case.
In particular, the maximal sensitivity (reached around $\delta_{13}$ $\sim$ $\pm$ $90^0$) 
decreases from  $8\sigma$ to $7\sigma$ if the amplitude of  the two new mixing angles 
$\theta_{14}$ and $\theta_{24}$ is close to that of $\theta_{13}$. 
The sensitivity to the CPV due to $\delta_{14}$ can reach 5$\sigma$ but
is below  3$\sigma$ for most of the true values of such a phase. 
We have also investigated the reconstruction capability of the two phases $\delta_{13}$ and $\delta_{14}$. 
The typical 1$\sigma$ uncertainty on $\delta_{13}$ ($\delta_{14}$) is $\sim15^0$ ($30^0$).
Finally, we have assessed the sensitivity to the octant of $\theta_{23}$. We have found
that in T2HK there can be a complete loss of sensitivity for unlucky values of the two CP phases 
$\delta_{13}$ and $\delta_{14}$. We highlighted, for the first time in the literature, the crucial role 
of the spectral energy shape information in improving the sensitivity both to the mass hierarchy and 
to CPV induced by the new CP-phase $\delta_{14}$. Notably, we have shown that events counting,
for unlucky values of the CP-phase $\delta_{13}$, may 
be completely insensitive to CPV induced by the new CP-phase $\delta_{14}$,
and that an appreciable sensitivity can be guaranteed only by the energy spectral information.
We hope that our present study to look for a light sterile neutrino in T2HK will
be a useful addition to the list of important physics topics which can be 
addressed using the proposed T2HK setup.

\subsubsection*{Acknowledgments}
S.K.A. and S.S.C. are supported by the DST/INSPIRE Research Grant [IFA-PH-12],
Department of Science \& Technology, India. A part of S.K.A.'s work was 
carried out at the International Centre for Theoretical Physics (ICTP), 
Trieste, Italy. It is a pleasure for him to thank the ICTP for the hospitality 
and support during his visit via SIMONS Associateship. S.K.A would
like to thank T. Kobayashi for useful discussions.
A.P. is supported by the grant ``Future In Research''  {\it Beyond three neutrino families}, 
Fondo di Sviluppo e Coesione 2007-2013, APQ Ricerca Regione Puglia, 
Italy, ``Programma regionale a sostegno della specializzazione intelligente 
e della sostenibilit\`a sociale ed ambientale''. A.P. acknowledges partial support 
by the research project {\em TAsP} funded by the 
Instituto Nazionale di Fisica Nucleare (INFN). 

\bibliographystyle{JHEP}
\bibliography{Sterile-References}

\end{document}